\pdfoutput=1
\documentclass{JINST}

\title{The NA62 LAV front-end electronics}

\author{A. Antonelli$^a$, G. Corradi$^a$, M. Moulson$^a$, C. Paglia$^a$,  M. Raggi$^a$\thanks{Corresponding author: M. Raggi} , T. Spadaro$^a$, D. Tagnani$^a$, 
F. Ambrosino$^b$, D. Di Filippo$^b$, P. Massarotti$^b$, M. Napolitano$^b$, G. Saracino$^b$,
B. Angelucci$^c$, F. Costantini$^c$, R. Fantechi$^c$, S. Gallorini$^c$, S. Giudici$^c$, I. Mannelli$^c$, F. Raffaelli$^c$, S. Venditti$^c$,
G. D'Agostini$^d$, E. Leonardi$^d$,  V. Palladino$^d$, M. Serra$^d$, P. Valente$^d$ \\
\llap{$^a$}INFN Laboratori Nazioni di Frascati,\\
  Via E. Fermi 40 Frascati, Italy\\
\llap{$^b$} Dipartimento di Scienze Fisiche Universtita' Federico II,\\
 via Cinthia,  Napoli, Italy\\
\llap{$^c$} Dipartimento di Fisica dell'Universita' and Sezione INFN,\\
   Largo B. Pontecorvo  3, Pisa, Italy\\
\llap{$^d$} Dipartimento di Fisica dell'Universita' and Sezione INFN Roma La Sapienza,\\
  Piazzale Aldo Moro 2 Roma, Italy\\

  E-mail: \email{mauro.raggi@lnf.infn.it}}

\abstract{The branching ratio for the decay $K^+\to \pi^+ \nu \bar{\nu}$ is sensitive to new physics; the NA62 experiment will measure it to within about 10\%. To reject the dominant background from channels with final state photons, the large-angle vetoes (LAVs) must detect particles with better than 1 ns time resolution and 10\% energy resolution over a very large energy range. Our custom readout board uses a time-over-threshold discriminator coupled to a TDC as a straightforward solution to satisfy these requirements. A prototype of the readout system was extensively tested together with the ANTI-A2 large angle veto module at CERN in summer 2010.}


\begin{document}

\section{The NA62 experiment at CERN}

The branching ratio (BR) for the decay $K^+\to \pi^+ \nu \bar{\nu}$
can be related to the value of the CKM matrix element $V_{td}$ with minimal theoretical uncertainty, providing a sensitive
probe of the flavor sector of the Standard Model. The measured value of the BR is $(1.73^{+1.15}_{-1.05})\cdot10^{-10}$
on the basis of seven detected events\cite{Artamonov:2009sz}.
NA62, an experiment at the CERN SPS, has  the goal of detecting $\sim$100  $K^+\to \pi^+ \nu \bar{\nu}$ decays with a S/B ratio of 10:1 \cite{NA62:TDR}.
The experiment will make use of a 75 GeV unseparated positive secondary beam. The total beam rate is 800 MHz, providing $\sim$50 MHz of $K^+$. The vacuum decay volume begins 102 m downstream of the production target and it's $\sim$100 m long. 5 MHz of kaon decays are observed in the 65-m long fiducial region. Large-angle photon vetoes are placed at 12 stations along the vacuum tank and provide full coverage for decay photons with 8.5 mrad < $\theta$ < 50 mrad. The last 35 m of the vacuum tube hosts a dipole spectrometer with four straw- tracker stations operated in vacuum. The NA48 liquid-krypton calorimeter \cite{Fanti:2007} is used to veto high-energy photons at small angle. Additional detectors further downstream extend the coverage of the photon veto system (e.g. for particles traveling in the beam pipe).
The experiment must be able to reject background from, e.g., $K^+ \to \pi^+\pi^0$ decays at the level of $10^{12}$. Kinematic cuts on the $K^+$ and $\pi^+$ tracks provide a factor of $10^4$ and ensure 40 GeV of electromagnetic energy in the photon vetoes; this energy must then be detected with an inefficiency of $\sim10^{-8}$. For the large-angle photon vetoes, the maximum tolerable detection inefficiency for photons with energies as low as 200 MeV is $10^{-4}$. In addition, the large-angle vetoes must have good energy and time resolution and must be compatible with operation in vacuum.

\subsection{The NA62 Large Angle Veto}
\begin{table}[htdp]
\caption{LAV stations description.}
\begin{center}
\begin{tabular}{|l|c|c|c|c|c|}
\hline
Station number & Inner diameter &  Outer diameter & \# layers & \# blocks & \# of RO ch\\
\hline
A1-A5   & 1073 mm & 1813 mm & 5 & 160 & 320\\
A6-A8   & 1534 mm & 2274 mm & 5 & 240 & 480\\
A9-A11 & 1960 mm & 2700 mm & 4 & 240 & 480\\
A12       & 2144 mm & 2854 mm & 4 & 256 &  512\\
\hline
total & & & & 2496 & 4992 \\
\hline
\end{tabular}
\end{center}
\label{Tab:LAV}
\end{table}

The large angle veto detector is made by 12 stations of different radii, ANTI-A1 to ANTI-A12, to assure an angular coverage for photons between 8-50 mrad.
The first eleven stations are part of the vacuum decay tube, while the last is located outside the vacuum tank.
There are 4 different types of veto station whose characteristics are listed in Tab \ref{Tab:LAV}.
Each type has a different number of layers, 4 or 5, of lead glass blocks, coming from the dismantled electromagnetic calorimeter of the OPAL experiment.
The layers are staggered in azimuth providing complete hermeticity of at least three blocks in the longitudinal direction.  
The blocks, made of Schott SF57 lead glass, have the shape of a truncated prism with different shapes and dimensions (with minimal variations between different types) . The block length is always 370 mm. One of the square faces of the lead glass has a 1 cm-thick steel flange glued to it. This flange has four threaded holes for fixing the counter to the support bracket, one for the connection of a calibration diode, and a central large hole for the passage of a cylindrical light guide for light collection. The light guide is a cylinder of SF57 lead glass with a diameter of 76 mm and a height of 40 mm. It is glued to the lead glass block and, at the other end, to a Hamamatsu R2238 photomultiplier. An external mu-metal shield, enclosing the guide and the PM, is glued to the steel flange.

\section{LAV readout}
The LAV system will mainly detect photons from kaon decays, as well as muons and pions in the beam halo. For each incoming particle the veto detectors are expected to provide a time measurement with ~1 ns resolution and an energy measurement with a moderate precision (of order 10\%). The system should be able to operate with thresholds of few millivolts, well below the minimum-ionizing-particle (MIP) signal, in order to keep the detection efficiency for muons and low energy photons as high as possible.
Because of the intrinsic time resolution of the lead-glass blocks (<1ns) and the rise time of the Hamamatsu R2238 PMT (~5 ns), the requirements are not stringent  on the time measurement accuracy. On the other hand, the expected energy deposit in the LAV stations from photons coming from $\pi^0$ decays covers a very wide range, from ~10 MeV up to 30 GeV.
Using the measured average photoelectron yield of 0.3 p.e./MeV and a nominal gain of $1\cdot10^6$ for the R2238 PMT, one expects a ~4.5 pC charge for a MIP, corresponding to a signal amplitude of ~20 mV on a  50$\Omega$ load. On the upper part of the range, signals from 20 GeV showers can reach an amplitude of 10V for a 50$\Omega$ load.
\begin{figure}[h]
\centering
\includegraphics[width=14.5cm]{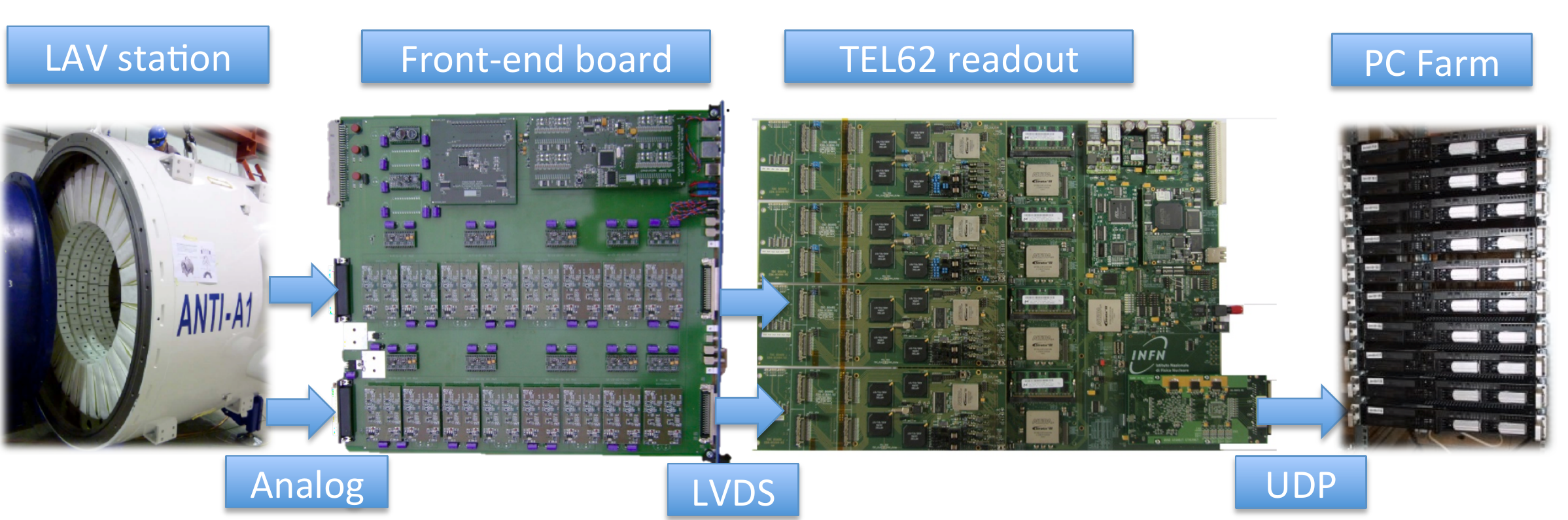}
\caption{\it The LAV readout scheme.}
\label{fig:lav_readout}
\end{figure}
The readout for the LAV stations consists of two different types of boards (see fig. \ref{fig:lav_readout}):
a dedicated front end card developed for the LAV detector, and a common digital readout board called TEL62, used by most of the NA62 detectors. 
The LAV front end board converts the analog input coming from the PMT into an LVDS digital signal. The duration of the LVDS pulse is 
equal to the time the analog signal is over a programmable threshold. 
The LVDS logic signal is sent to the readout board TEL62 in which a custom designed TDC mezzanine converts the signal in digital leading and trailing times.
The TEL62 on-board FPGAs are used to correct raw hits times and to produce a L0 trigger primitive to be sent to the L0 trigger processor. On a positive L0 trigger request the TEL62 sends the data to the L1 PCs using its dedicated 4x1Gbit ethernet interface. The system has $\sim2500$ analog input channels and $\sim5000$ digital readout channels in total.

\section{The LAV front end electronic board}

\begin{figure}
\centering
\includegraphics[width=8.cm]{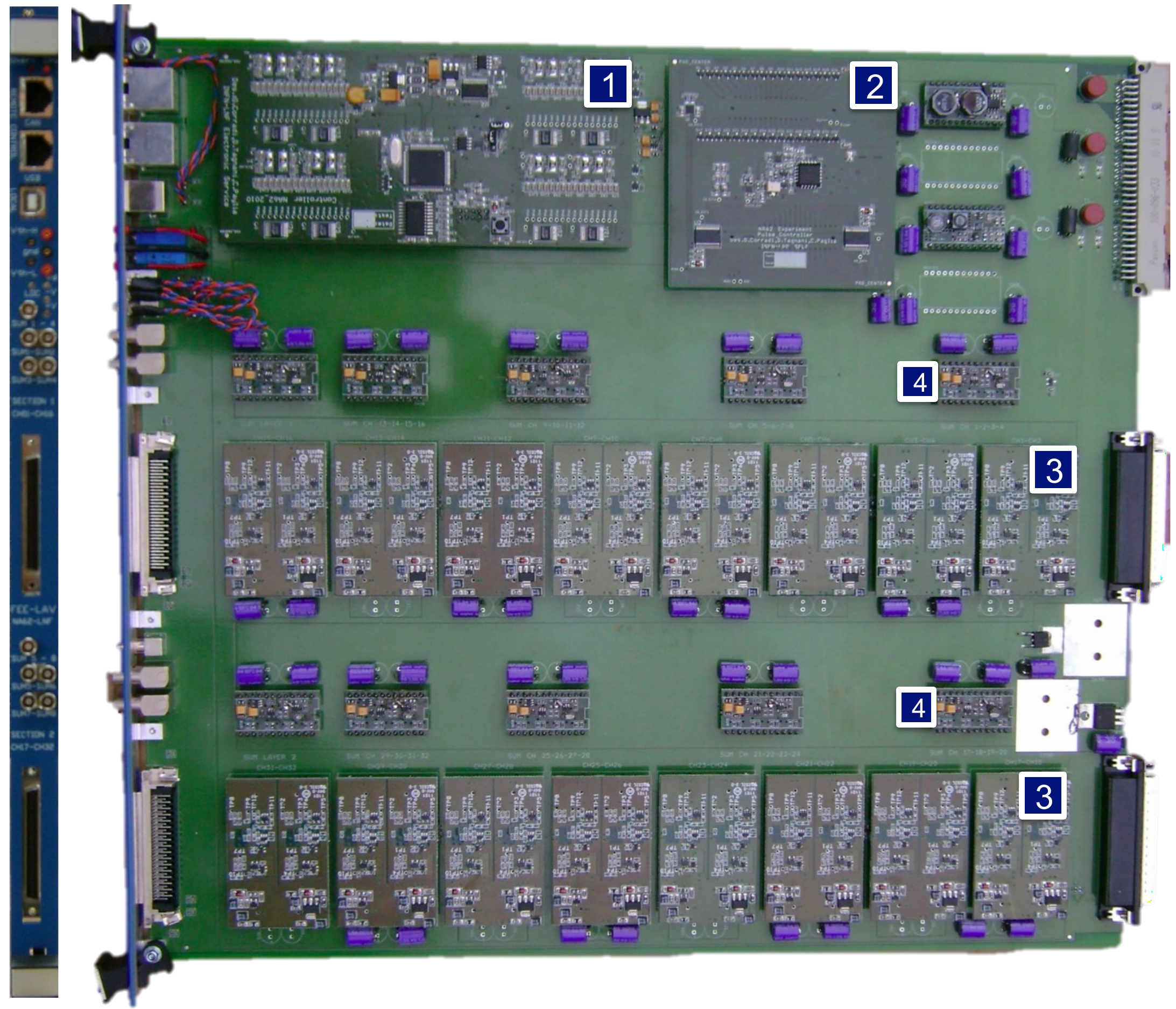}
\vspace{-0.3cm}
\caption{\it The LAV front end board.}
\label{fig:LAV9U}
\end{figure}

The LAV front end board is implemented on a 9U VME standard layout with the J1 power connector only at the top of the backplane side. At the bottom the 32 analog inputs are connected to the board using two DB37 connectors (see right side of fig \ref{fig:LAV9U}). The power to the J1 connector is provided by non-VME-standard pins at $\pm12V$,  reduced to $\pm7V$ by custom designed very low noise switching voltage regulators, and then distributed along all the board. Each single input produces two different outputs due to the presence of two programmable thresholds on each channel. The resulting 64 LVDS digital outputs are connected to the TDC using two SCSI2 connectors. The analog sums of 4 and 16 channels are provided on 8 + 2 LEMO00 connectors for monitoring of the analog signal. The communication and the threshold setting are managed by the CANOPEN protocol through two RJ-45 connectors. To simplify maintenance and reduce costs the board has a modular structure. The 9U motherboard manages input, output and power distribution while the rest of the functionalities are implemented on 4 types of mezzanine described in following sections:

\begin{enumerate}

\item Board controller mezzanine
\item Test pulse generator mezzanine
\item Time over threshold (ToT) mezzanine
\item Sum of 4 mezzanine
\end{enumerate}

%The functionality of each of the mezzanines is described in following sections.

\subsection{Board controller}
This mezzanine is responsible for the communication with the slow control PC and for the setting of the 64 comparators threshold. 
 It is based on the 80F40 CPU. It communicates through the CANOPEN protocol and allows different operations to be performed. The slow control
 functionality includes the monitoring of the low voltage power lines and the measurement of the board temperature. The board also communicates with
 the ToT mezzanine allowing to set and read the thresholds on each comparator, and with the pulse generator mezzanine allowing to set:
 the value of the pulse height and frequency and the pattern of channels to be pulsed.

\subsection{Test Pulse Generator}
For use in the diagnosis of the functionality of the electronics, channel mapping, channel integrity and threshold calibration, the design of the board includes an internal pulse generator. 
The pulse generator is able to provide pulses with 5-50 ns programmable width and 50-500 mV programmable amplitude using 12 bit words. 
The stability of both amplitude and width is of the order of 3\% and the pulse rise time will be  $\sim$2 ns. The internal pulser can be triggered both locally,
 controlled by the on board CPU, or externally using an external clock. To allow maximum flexibility, the pattern of channels to be pulsed can be programmed.
 The pulse signal is directly sent to the input connectors. Exploiting the high impedance of the signal input line from the PMT voltage divider, as well as the fact that 
 the pulser signal travels both in the detector direction and the comparator direction, by measuring the time distance of the two pulses the integrity of all the electronic chain up to the Tel62 can be checked.

\subsection{ToT mezzanine}
This is the core part of the LAV FEE which manipulates the analog signal to produce the LVDS output. 
The LAV FEE board houses 16 of these mezzanines which are able to manage 2 input channels each.
Just after the input connection the signal is divided into two copies using a passive resistive splitter.
One copy is sent to the circuit responsible for the sums while the other is sent to the ToT chain.
In order to avoid channel saturation by the input signals, the input amplitude must be limited to a maximum of 600 mV.
The clamping circuit is designed to be able to sustain high rate of signal up to 10 V, but is able to tolerate even larger isolated signals. 
To preserve the ToT measurement, the circuit must clamp the signal without changing its time duration. This is achieved by an active clamp using a pair of very fast low capacitance diodes and the amplifier. The ToT system must work with an effective threshold, of a few mV on the analog signal in order to maximize efficiency. To improve signal to noise separation and reduce the walk dependence on the analog amplitude (overdrive), a moderate amplification is needed. A gain of 3 was chosen. 
A vey low noise, high bandwidth (800MHz), high speed Current Feedback Amplifier (type AD8001) is used for this purpose. After a decoupling capacitor, to suppress amplifier DC offset the output is sent to the comparator input.
\begin{figure}[h]
\centering
\includegraphics[width=12.5cm]{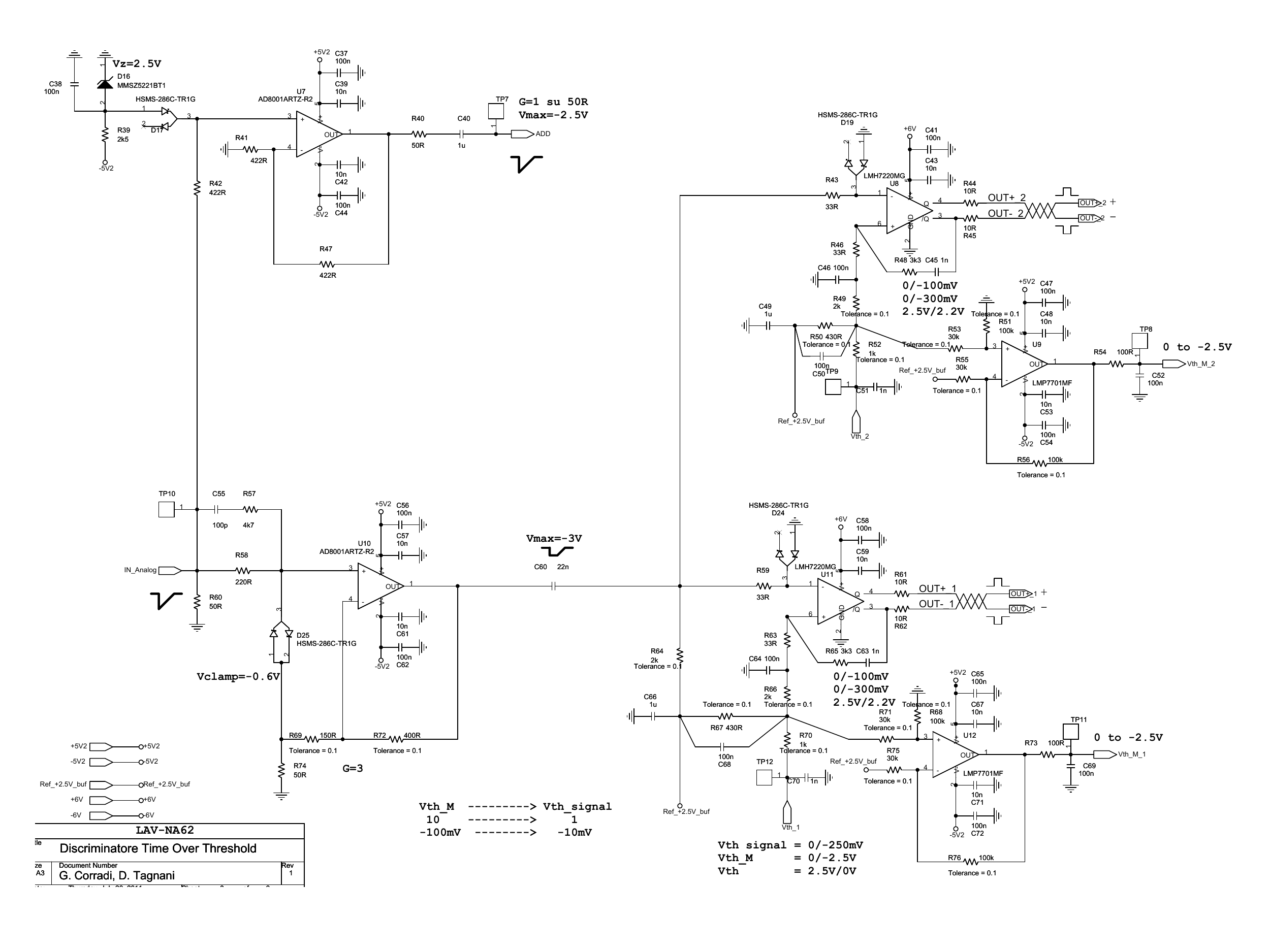}
\vspace{-0.3cm}
\caption{\it Time over threshold circuit single channe layout.}
\label{fig:ToT_lay}
\end{figure}
The amplified signal is picked up at high impedance by two LMH7220 High Speed Comparators with LVDS drivers. These devices compare the input with programmable thresholds which can be adjusted in the range 5-250 mV with a 12 bit resolution using the DAC in the board controller mezzanine. 
The LMH7220 has only a 2.5 ns propagation delay and 0.6 ns rise and fall times on the LVDS signal, minimizing the impact on the TDC performance.
To reduce noise-induced double pulses at the comparator output, a $\sim$3 mV hysteresis is also provided through a feedback resistor. The comparator produces an output signal starting when the leading edge of the analog signal crosses the threshold, and stopping when the trailing edge crosses the threshold again. This digital signal is transmitted to the TDC using the LVDS differential standard. To dump the effect of impedance mismatch a pair of output resistors are used on the LVDS lines.

\subsection{Analog Sum mezzanine}
\begin{figure}[h]
\centering
\includegraphics[width=10.cm]{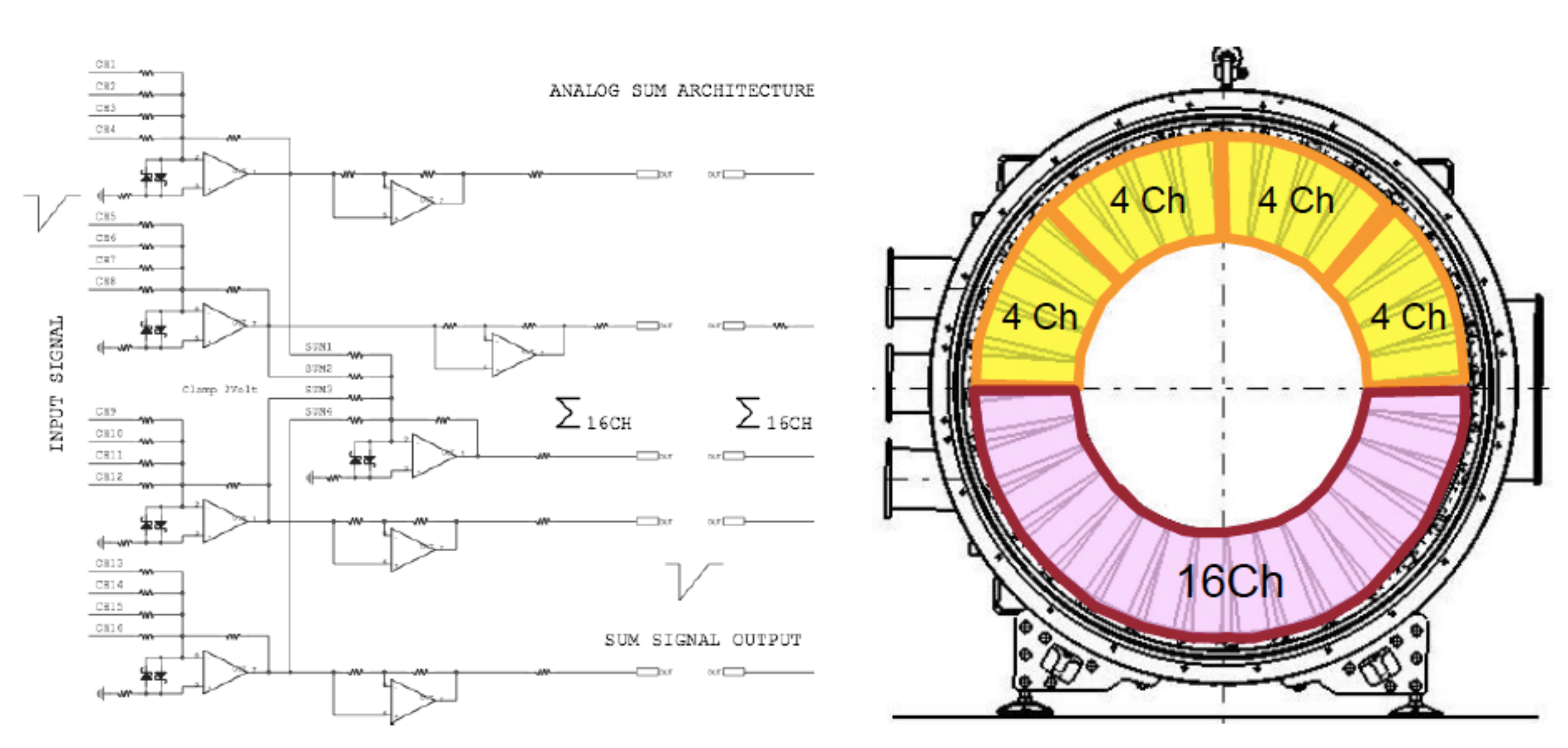}
\vspace{-0.3cm}
\caption{\it Analog sums layout. On the right side the geometry of sums on the smaller ANTI, e.g. A1 to A5.}
\label{fig:sums}
\end{figure}

In order to have the possibility to monitor the input analog signals to the FEE board, an analog output is required. 
Due to mechanical constraints it is impossible to duplicate all 32 analog inputs on LEMO connectors on the front-panel. 
The analog input signals are therefore collected in sums of four blocks (one azimuth segment) and then summed again in groups of four to get a sum of 16 blocks. 
The inputs to each sum are clamped at 600 mV before the summation is performed (see Fig. \ref{fig:sums}). The FEE board front panel is equipped with 
8 LEMO connectors for the sums of 4 blocks, (in yellow in fig. \ref{fig:sums}), and two LEMO connectors for the sums of a 16 blocks, (in violet in fig. \ref{fig:sums}). 
Using a digitizer, the total charge on 4 or 16 channels can be measured.

\section{Performance}
\begin{figure}[h]
\centering
\includegraphics[width=13.cm]{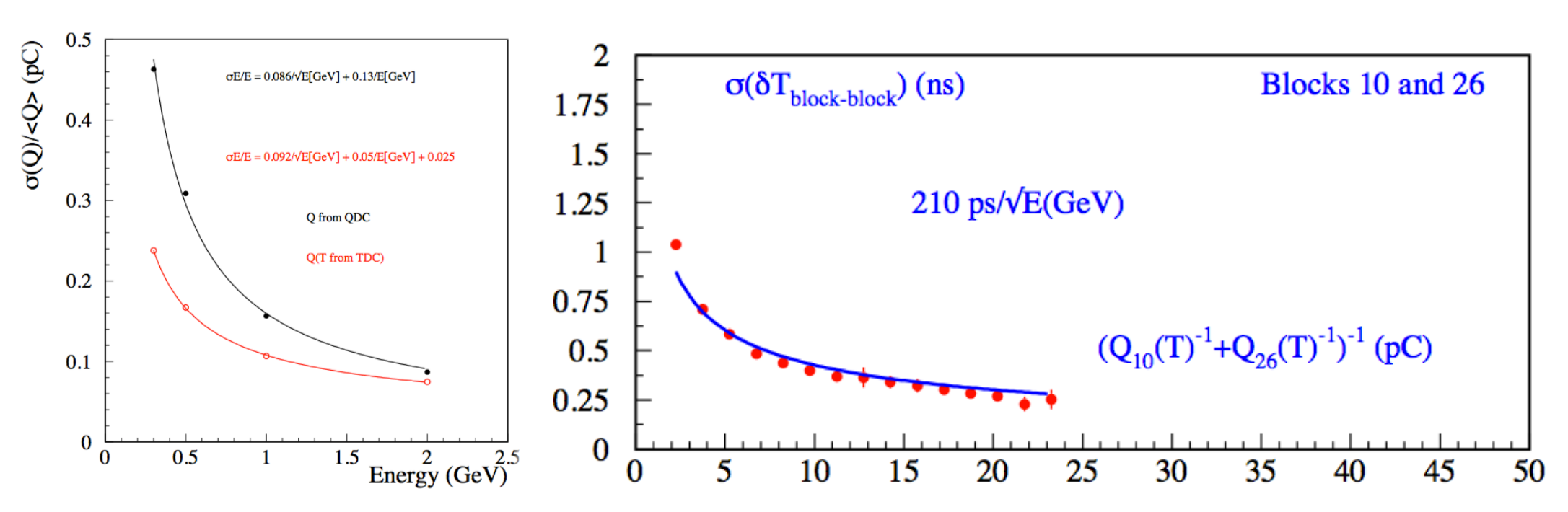}
\vspace{-0.3cm}
\caption{\it Energy and time resolution from 2010 test beam.}
\label{fig:perfo}
\end{figure}

The performance of the LAV front end boards was extensively studied during the LAV test beam at CERN in fall 2010 at the T9 beam line. Due to the unavailability
 of the 9U boards the ToT and the sum circuits were tested using 16 ch VME 6U prototypes. For that purpose, 6 prototype boards for a total of 96 channels
 were produced and used in the test beams. The readout was based on VME standard electronics and used two parallel solutions:
 the charge measurement was performed using a commercial CAEN V792 QDCs, and the ToT measurement was performed using commercial CAEN V1190B TDC. 
 The time resolution, energy resolution and effective threshold were studied see fig. \ref{fig:perfo}.
The measurement of the energy resolution is based on a sample of pure electrons selected using threshold Cherenkov detectors on the beam line. The two different curves represent the resolution obtained with different measurement techniques. The black curve is obtained using the charge measured by the QDC, while the red curve is obtained by reconstructing the charge from the ToT measured by the TDC. At low energies the ToT technique gives better results than the direct charge measurement. This is due to the very steep dependence of ToT on the charge in the low energy region.  The time resolution is obtained using the difference of the hit time of two blocks on first and second ring of the LAV.

\section{Simulation}
\begin{figure}[h]
\centering
\includegraphics[width=7.5cm]{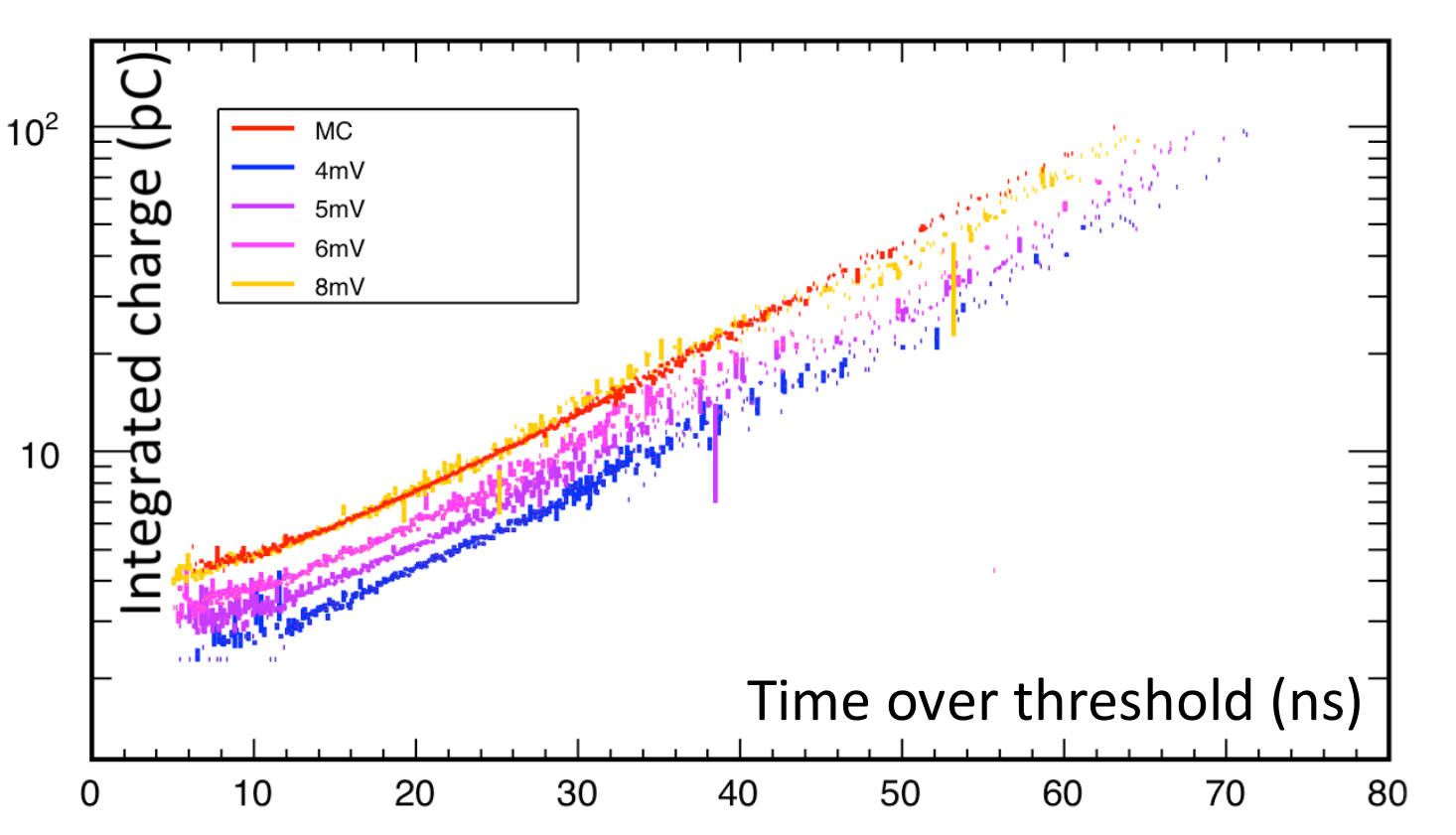}
\vspace{-0.3cm}
\caption{\it Time over threshold vs charge.}
\label{fig:simul}
\end{figure}

In order to better understand the detector and the electronic performance a detailed MC simulation has been developed for the digitization procedure.
In the simulation, the number of photons on the PMT photocathode their energies and arrival times are obtained from GEANT4. The behavior of the R2238 PMT is then simulated including the photocathode quantum efficiency as a function of the wavelength, multiplication process in the 12 dynodes, the output RC 
circuit, and the cables. The simulated PMT signal is then processed by the simulation of the front end which, taking into account thresholds and hysteresis, produces the value of the simulated leading and trailing times. To validate the simulation, the curve of time over threshold vs charge has been compared with test beam data. The result is shown in figure \ref{fig:simul}. Very good agreement with the data is obtained over a large range of charge from a few pC to ~100 pC.

\section{Conclusions}
A low cost, large dynamic range, Time-over-threshold based solution has been developed for the LAV front end electronics. The achieved time resolution 
has been measured to be of the order of $200ps/\sqrt E$ while the energy resolution $\sigma(Q)/Q=9.2\% / \sqrt E \oplus 5\% /E \oplus 2.5\%$. 
Both characteristics are dominated by the detector performance and fulfill the NA62 LAV requirements.

%\acknowledgments

\end{document}